\documentclass[oldversion]{aa}
\usepackage{graphicx}
\usepackage[]{natbib}
\usepackage{mathrsfs}
\usepackage{amsmath}
\usepackage{amssymb}
\bibpunct{(}{)}{;}{a}{}{,}

\newcommand{\chan}[0]{\textit{Chandra}}
\newcommand{\xmm}[0]{XMM-\textit{Newton}}
\newcommand{\au}[0]{\mbox{AU~Mic}~}

\setcitestyle{notesep={ }}

\begin{document}

\title{X-raying the AU Microscopii debris disk}
\titlerunning{X-raying AU Mic's disk}
\author{P. C. Schneider\and J. H. M. M. Schmitt}
\institute{Hamburger Sternwarte, Universit\"at Hamburg, Gojenbergsweg 112, 21029 Hamburg, Germany\\
\email{cschneider@hs.uni-hamburg.de}}
\date{Received ... / accepted ...}
\abstract
{
AU~Mic is a young, nearby X-ray active M-dwarf with an edge-on debris disk. Debris disk are the successors of the gaseous disks usually surrounding pre-main sequence stars which form after the first few Myrs of their host stars' lifetime, when -- presumably -- also the planet formation takes place.  Since X-ray transmission spectroscopy is sensitive to the chemical composition of the absorber, features in the stellar spectrum of AU~Mic caused by its debris disk can in principle be detected. The upper limits we derive from our high resolution {\it Chandra}~LETGS X-ray spectroscopy are on the same order as those from UV absorption measurements, consistent with the idea that AU~Mic's debris disk possesses an inner hole with only a very low density of sub-micron sized grains or gas.
}

\keywords{Stars: circum-stellar matter, Stars: individual: AU Mic, X-rays: stars }
\maketitle
\section{Introduction}

The disks around young stars undergo dramatic changes during the first $\sim10$~Myrs after their host stars' birth, when the gas content of the disk largely disappears \citep{Alexander_2008, Meyer_2007, Hernandez_2006}, leaving behind a so-called debris disk. The Kuiper belt and the asteroid belt are the solar system's analogs of stellar debris disks. The main components of an optically thin debris disk are small grains with about sub-micrometer sizes, larger bodies in the cm range and, possibly, planets, which are thought to form in the same time-span.
The initial composition of the material in the debris disks after the transition phase is not well known.  Collisions of already formed smaller bodies replenish the dust in the ``older'' debris disks, while it is not clear, whether this is also the source of the initial dust in the debris disk or whether it is remnant proto-planetary dust.

\subsection{AU Mic and its activity}
\au is a $12^{+8}_{-4}$~Myr old M1 dwarf at a distance of about 10~pc,  which belongs to the $\beta$~Pic moving group \citep[e.g.][]{Navascues_1999, Zuckerman_2001}. \au is one of the brightest nearby X-ray emitters ($\log L_X \approx29.3$) and shows strong flaring activity, making \au a valuable target for flare studies as shown by e.g. UV observations \citep{Robinson_2001}.

At X-ray wavelengths \au has been observed many times. The first \chan~ observation provided the highest resolution spectrum, but was limited to the wavelength range below $\sim 25$~\AA~ \citep{Linsky_2002}.
The \xmm~ observation of \au in 2000 was simultaneous with UV and VLA observations revealing several flares  \citep[][for line fluxes]{Smith_2005,Mitra-Kraev_2005,Ness_2003}.
Furthermore, \au was the target of FUSE and Hubble Space Telescope STIS observations \citep[e.g.][]{Pagano_2000, Robinson_2001, DelZanna_2002}, aiming at the determination of the temperature structure of its chromosphere and corona.

\subsection{AU~Mic and its debris disk}
The first indications for cold material around \au go back to IRAS data,
 which exhibit excess emission at 60~$\mu$m \citep{Mathioudakis_1991,Song_2002}.
A clear infrared excess at 850~$\mu$m in the spectral energy distribution (SED) of \au was detected by \citet{Liu_2004}, clearly pointing to the existence of a debris disk. By assuming an optical thin disk at a single temperature, \citet{Liu_2004} derived a mass of 0.01~$M_\oplus$ and a temperature of 40~K for the disk. These values have been confirmed by \citet{Rebull_2008} from Spitzer data \citep[see also][]{Chen_2005}.
\citet{Metchev_2005} also confirmed the dust mass of about 0.01~$M_\oplus$ composed of grains in the submicron regime by modelling the optical, near-IR and SED data.

Optical observations by \citet{Kalas_2004}, initiated shortly after \citet{Liu_2004}, clearly showed the presence of a debris disk around \au with a radial extent of at least 210~AU and almost perfectly edge-on.
The disk has then been subsequently studied at optical wavelengths with, e.g., the Hubble Space Telescope (HST) and adaptive optics, making it one of the most well studied debris disks.
The models derived by \citet{Krist_2005} from their HST observation restrict the disk inclination to $i \gtrsim 89^\circ$; they also confirmed the small-scale brightness variations detected by \citet{Kalas_2004}. 
These disk inhomogeneities can be readily explained by the existence of orbiting planets, however, no clear signatures of a planet have been found to date \citep[][]{Hebb_2007, Metchev_2005}.

Two studies aimed at the detection of circum-stellar gas in the \au disk. By their non-detection of far-UV H$_2$ absorption, \citet{Roberge_2005} derived an upper limit on the gas column density along the line of sight of $N_{H_2} < 10^{19}$~cm$^{-2}$. The detection of $H_2$ in fluorescence enabled \citet{France_2007} to derive a column density of  $3\times10^{15}\, \text{cm}^{-2} < N_{H_2} < 2\times10^{17}\, \text{cm}^{-2}$ ($T_{H_2}=800$~K and 2000~K, respectively). Comparing their $H_2$ value with the the CO results of \citet{Liu_2004}, they conclude that \mbox{$H_2$} contributes less than about 1/30$^{th}$ to the total disk mass.
The $H$ absorption mainly traces interstellar rather than circum-stellar material and has a column of $N_H = 2.3\times10^{18}$~cm$^{-2}$ \citep{Wood_2005}, thus corresponding to within a factor of five to the Mg~{\sc ii} absorption measurements of \citet[][, $N_{Mg} = 1.6\times10^{13}$~cm$^{-2}$]{Redfield_2002}, assuming solar abundances and Mg~{\sc II} to be the dominant Mg species \citep{Slavin_2002}.

\subsection{Disk models}
\citet{Krist_2005} used three-dimensional models of the scattering cross-section densities throughout the disk to 
interpret their HST images. They find that
in the inner disk region \mbox{(12~AU $<r<$ 49~AU)} the observations can be explained by a relatively uniform scattering cross-section density (forward scattering particles with an albedo of 0.5) in approximate correlation with the non-flaring disk model of \citet{Metchev_2005}.
\citet{Augereau_2006} inverted the visible and near-IR scattered light profiles to study the grain properties and find that the scattered light is reflected at grains with sizes mainly between 0.1~$\mu$m and 1~$\mu$m. However, they require only about 1/20$^{th}$ of the total disk mass to account for the scattered light.

The dynamics of the grains were included by \citet{Strubbe_2006} into their models and led them to
explain the observations with a ``birth ring'' at about 40~AU, where larger planetesimals of about decimetre size are located. By collisions of these planetesimals the submicron sized grains are produced, which then, depending on their size, the radiation pressure, the gas content of the disk and the stellar wind, are expelled from the system or dragged into AU~Mic.
\citet{Strubbe_2006} conclude that AU~Mic's debris disk is dominated by destructive grain-grain collisions and that the inner part of the disk is largely void of submicron sized grains. These small grains mainly populate the outer part of the disk and result in the blueish appearance of a debris disk relative to the star, since submicron sized grains provide the largest fraction of the scattered light \citep{Metchev_2005}. In contrast to the scattered light, the IR excess is caused by larger bodies because the mass of the submicron sized grains is too low to account for the observed emission \citep{Fitzgerald_2007}. \citet{Augereau_2006} already noticed the need to increase their disk mass derived from the scattered light to reproduce the SED.
The void of small grains in the inner part of the disk is consistent with the polarisation data of \citet{Graham_2007} and the near-IR data of \citet{Fitzgerald_2007}, who derived an upper limit on the mass of submicron sized grains in the inner zone of the disk of $10^{-4}M_\oplus$.

\begin{figure}%[ht!]
\centering
  \includegraphics[width=0.49\textwidth]{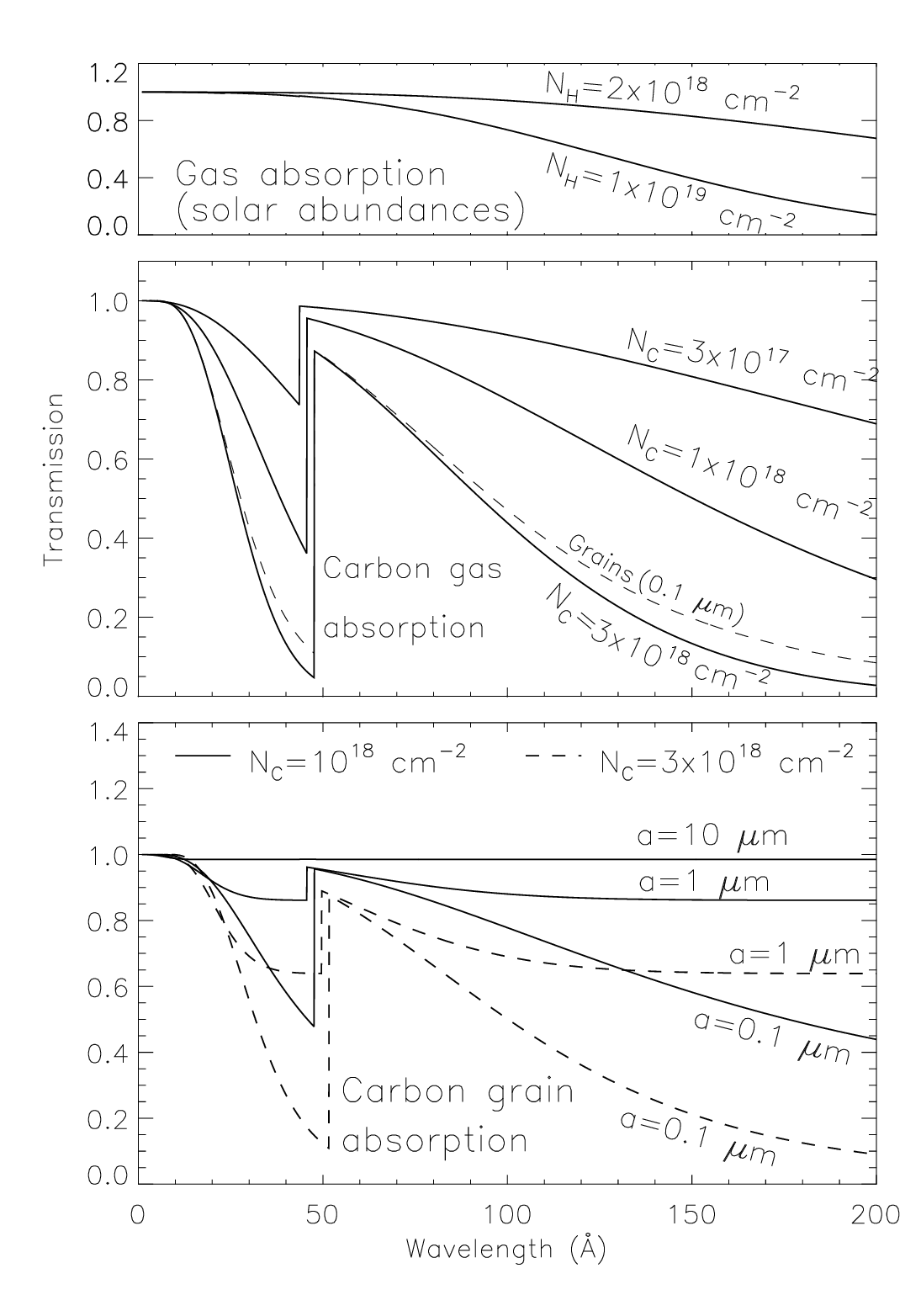}
  \caption{X-ray transmission curves for different absorbers using \texttt{bamabs} from the PINTofALE-package based on the 
data of \citet{Balucinska_1992}. 
\textbf{Top}: Absorption by cold gas with solar abundances. \textbf{Middle panel}: Gaseous carbon absorption. \textbf{Bottom}: Absorption by carbon grains. The individual curves are slightly shifted along the x-direction to maintain clarity. The C-K edge is located at $43.6$~\AA.\newline
  For the grain absorption, the column density gives the total number of carbon atoms along the line of sight. Therefore, the number of grains in the line of sight depends on the grain-size, i.e., varies between the transmission curves for a fixed column density.
  For comparison, the dashed line in the gas absorption panel describes the transmission of carbon grains with $N_C = 3\times10^{18}\,\mbox{cm}^{-2}$.  \label{fig:C_trans}}
\end{figure}

\section{The role of X-rays\label{sect:Xabs}}

The disk models strongly suggest that AU~Mic is directly shining through its disk, therefore, absorption features related to the disk should in principle be present in the observed spectrum. \au 's strong X-ray emission make this system a prime target to search for X-ray absorption features from a circum-stellar disk. The dominance of dust lets us expect a substantial amount of carbon in \au 's disk, and thus we examine the influence of carbon on an X-ray transmission spectrum.

\subsection{X-ray transmission in the disk of \au }
X-ray transmission spectra contain absorption features directly related to  individual elements and, therefore, can be used to probe the elemental composition of an absorber \citep[e.g.][]{Nicastro_2005,Williams_2007}.
{\bf In particular, X-ray transmission spectroscopy allows an assessment of the carbon column density in the disk of \au}. 
In Fig.~\ref{fig:C_trans} we therefore illustrate the effects of X-ray absorption on the transmission curves for different column densities
and the resulting soft X-ray spectrum near the carbon K-edge and out
to 200 \AA\  , i.e., in the band pass exclusively accessible to the {\it Chandra} LETGS.  We specifically consider the case of cold
gas absorption with solar abundances (Fig.~\ref{fig:C_trans}, top panel), the case of a pure carbon absorber with various column 
densities (Fig.~\ref{fig:C_trans}, middle panel), and the case of carbon absorbers locked up 
in various grain sizes (Fig.~\ref{fig:C_trans}, bottom panel); clearly, 
for an absorber significantly composed of grains, self-shielding within the grains becomes important for the calculation of its transmission properties \citep[see Appendix~A in][]{Wilms_2000}.
From Fig.~\ref{fig:C_trans} (bottom panel) it is clear that graphite grains with sizes around 0.1$\,\mu$m are ideal to derive column densities from the carbon absorption edge, because they are so small that self-shielding is unimportant; the derived column density represents (almost) the number column density of carbon. With increasing grain sizes, the absorption features approach that of grey absorbers and grains with sizes in excess of $10\,\mu$m are virtually completely grey at X-ray wavelengths with only a marginal reduction of transmission.
Therefore, the correspondence of absorption edge depth and column density holds strictly only for gas and small grains, while it breaks down for unknown grain sizes (in particular for sizes $\gtrsim 0.3$~$\mu$m).

Fig.~\ref{fig:C_trans} also illustrates that it is virtually impossible to reconstruct the elemental composition of an absorber from the reduced transmission outside the absorption edges. The shape of the reduced transmission at longer wavelengths can be mimicked by choosing appropriate column densities for almost every composition, e.g., the differences in transmission between a pure oxygen or a pure helium 
absorber in the \chan ~LETGS wavelength range longwards of 50~\AA~  is less than 10~\%.
Note that only very few narrow absorption lines are available in the case of (low ionised) disks. In particular the 1s-2p transitions of oxygen (23.5~\AA) and carbon ($\sim$44.8~\AA, estimated from the K$\alpha$ energy) are available \citep{Wilms_2000, Henke_1993}.

In the case of an absorber with solar abundances the absorption due to H, He and O reduces strongly the transmission around the absorption edge of C, however, the flux ratio between both sides of the edge depends only on the actual carbon column density.  For a column
density of $N_H = 10^{21}$~cm$^{-2}$ assuming cold gas with solar abundances, the transmission is reduced by a factor of 20, but
the transmission at the low energy side of the C-K edge is about twice that of the high energy side, yet 
the resulting low flux level makes it virtually impossible to measure the depth of the C-K edge. 
Therefore, using the carbon absorption edge requires carbon to be strongly enhanced in the absorber so that it contributes significantly to the absorption without the overall reduced transmission by other elements.

\subsection{The mass column density of \au}

The range of possible disk masses and hence column densities for AU~Mic is relatively narrow. The different methods 
independently point to a total disk mass of $0.01\,M_\oplus$ \citep[$\approx 6\times10^{25}$~g, cf., Sect. 4.1 in ][]{France_2007, Fitzgerald_2007}, however, the distribution of this disk mass between small grains and larger bodies and their locations are not particularly well 
constrained. The most recent disk models including the grain dynamics attribute less than a tenth of the total mass to small grains, while most of the mass is stored in larger bodies with sizes up to $\sim$~10~cm.  In order
to test these scenarios we calculate the mass column density by distributing 0.01~$M_\oplus$ uniformly within the inner zone of the \citet{Krist_2005} model (10~AU -- 50~AU, $h\approx 2$~AU), i.e., the zone where the dust causing the infrared excess should be located. The mass column density $\sigma$ is given by
\begin{eqnarray}
\sigma = \frac{M}{V}d &=& \frac{M}{\pi h(r_o^2-r_i^2)}\cdot (r_o - r_i) = \frac{M}{\pi h(r_o+r_i)}\\
 & \approx & 7\times10^{-4} \text{g}/\text{cm}^2\, \nonumber,
\end{eqnarray}
where $M$ is total disk mass within \mbox{50 AU}, $V$ is the volume occupied by the disk, $d$ is the line of sight and $r_o$ and $r_i$ are the outer and inner radii of the considered region, assuming no significant contribution of larger bodies.

As discussed above, the associated X-ray absorption features depend on the chemical composition of the disk and the grain sizes and shapes.
Although \citet{Roberge_2006} found that in the \mbox{$\beta$ Pic} disk the chemical composition might deviate from solar by factors of a few with carbon being overabundant, we assume for the \au disk that H and He are virtually absent and that the mass is stored in the remaining elements with solar abundances. This implies that $\sim 15$\% of the mass is stored in carbon atoms. The number density of carbon atoms in the line of sight is then
\begin{equation}
N_C = X_C\frac{\sigma}{m_C}  \approx 0.15\frac{7\times 10^{-4}\text{g}/\text{cm}^2}{12 \cdot 1.66\times 10^{-24}\text{g}} = 5\times 10^{18}\text{cm}^{-2},
\end{equation}
where $X_C$ is the mass-fraction of carbon (15~\%) and $m_C$ is the mass of a carbon atom.

From Fig.~\ref{fig:C_trans} it is clear that such column densities should impose strong feature on the transmission spectrum detectable with a \chan~ LETGS observation.

\begin{figure}[t]
  \includegraphics[width=0.49\textwidth, angle=0]{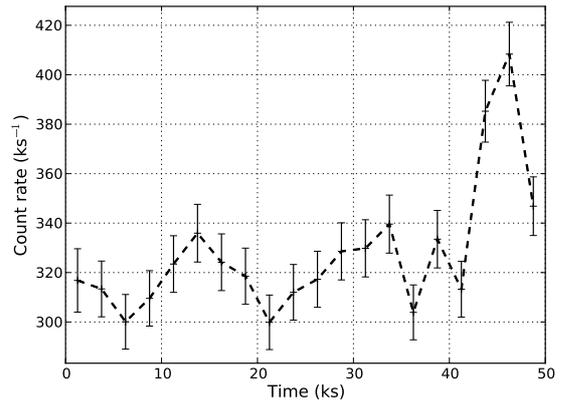}
  \caption{LETGS zeroth order X-ray lightcurve of AU~Mic. Time represents the time after the start of the observation (MJD=54643.506).\label{fig:flare}}
\end{figure}

\section{Observations, data reduction and immediate results\label{sect:iresults}}

\subsection{Observation and data reduction}

Au~Mic was observed on 30 June 2008 with the \chan~LETGS (Obs-ID 8894). The total exposure time was 50~ksec  and the data reduction was carried out using CIAO~4 \citep{CIAO}.
The photon detector of the LETGS is the HRC-S, a micro-channel plate, which does not provide sufficient energy resolution to separate the individual diffraction orders superposed on the same detector area. Therefore, 
to allow an analysis of the data with standard tools like XSPEC \citep{XSPEC}, we constructed new response matrices including up to ninth diffraction order contributions  following the instructions given in the CIAO threads\footnote{\texttt{http://cxc.harvard.edu/ciao/threads/hrcsletg\_orders/}} (see Appendix~\ref{sect:RespHigherOrders}).
Line fluxes were obtained using \texttt{CORA} which accounts for the Poisson character of the data \citep{Ness_2002c}.

We experimented with the ``standard'' filters (light, medium and heavy) and with the new \texttt{Gain Map and Pulse-Height} filter\footnote{\texttt{http://cxc.harvard.edu/contrib/letg/GainFilter/}}. Around the carbon edge, the difference in the background fraction is only a few percent between the light and medium filter while significant for the new filter; we list the corresponding values in Table~\ref{tab:edge_contr}.
The figures and numbers given in the text pertain to the standard light filter, unless otherwise noted.

\begin{figure*}
\centering
  \includegraphics[width=0.95\textwidth, angle=0]{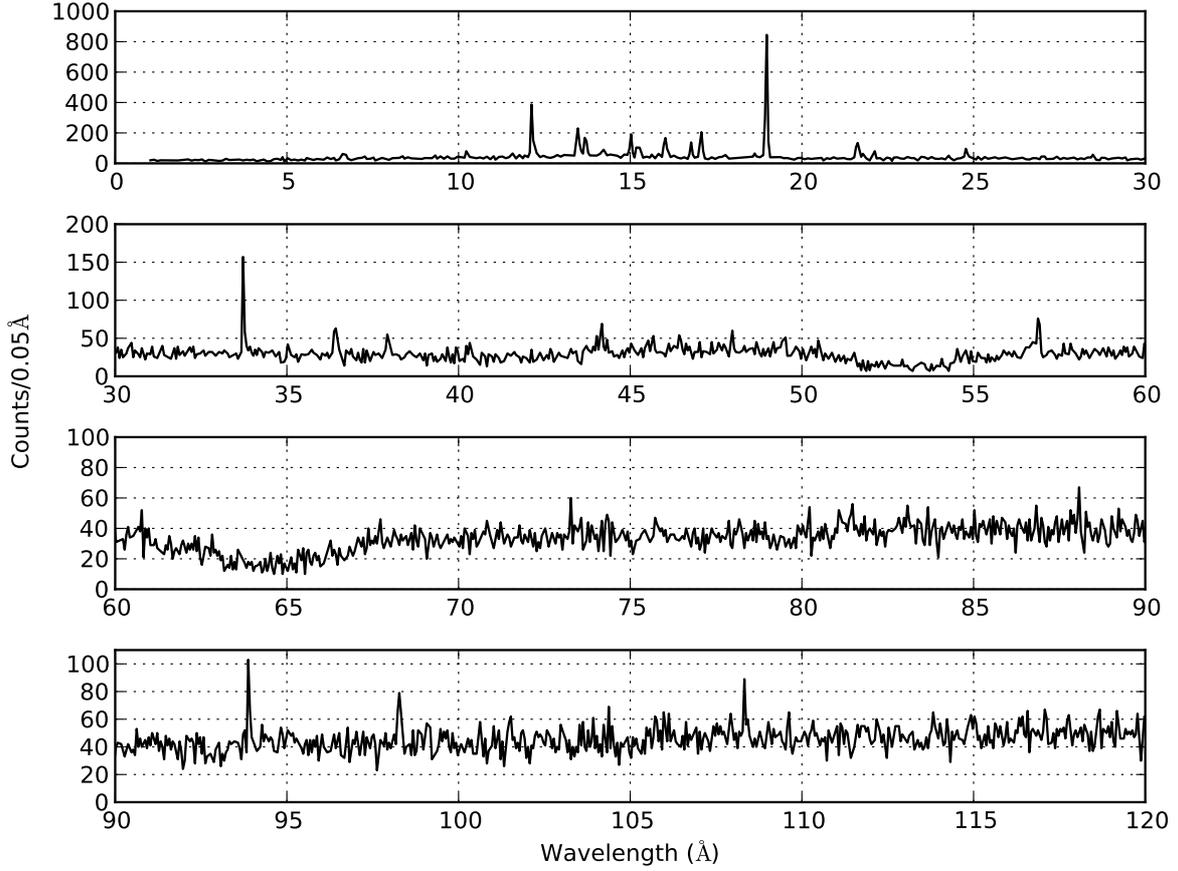}
  \caption{The measured spectrum in the spectral range relevant for the analysis. Shown is the sum of the positive and the negative diffraction order. \label{fig:completeSpectrum}}
\end{figure*}

\subsection{Global plasma properties}
In Fig.~\ref{fig:flare} we show the zeroth order X-ray count rate of AU~Mic vs. time; obviously, the light curve is more or less constant
during the first 40 ksec of the observations, afterwards there is a small flare-like increase.  We do not treat this increase separately
but analyse the observation in total.
In Fig.~\ref{fig:completeSpectrum} we show the recorded LETGS spectrum up to 120~\AA ; sub-regions will be shown individually in the next sections.  \au 's X-ray spectrum is typical of an active star, showing the strong Ne emission features and a strong OVIII Ly$_{\alpha}$ 
line; the carbon Ly$_{\alpha}$ line is also quite strong. The OVII triplet between 21.6 and 22.1 
\AA \  as well the Fe XVII feature at 15.03 \AA \ are relatively weak as well as all features attributable to N; we do point out the emission lines between 90 and 120 \AA \ attributable to highly ionised iron.  We used
XSPEC to fit the X-ray spectrum using a combination of three APED models \citep[variable abundances,][]{APED} with one absorption component. 
The increasing background at longer wavelengths make these bins less useful for the spectral analysis. We therefore restrict the wavelength range to values between 5~\AA~ and 35~\AA. Using the full wavelength range does not influence the two low temperature components; only the best fit temperature of the high temperature component doubles. This is mainly driven by a single, somewhat strangely shaped {Fe \sc XX} line at 132.85~\AA. Furthermore, the abundances of Mg, Al, Ca and Ni (low FIP) were coupled to that of Fe in order to decrease the number of free parameters. The thus obtained fit results (listed in Tab.~\ref{tab:XSPEC}) show the inverse FIP effect usually found in M-dwarfs \citep{FIP}.
The temperature structure and the abundances compare well with a fit performed with the \xmm ~RGS-data of AU~Mic \mbox{(Obs-ID 0111420101)}.

\begin{table}
\caption{Coronal properties of \au. Abundances are given relative to that of \citet{Asplund_2009} and 90~\% errors are given. \label{tab:XSPEC}}
\begin{center}
\begin{tabular}{cc}
\hline
\hline
Property & Value \\
\hline
$N_H$ & $2\times10^{18}$~cm$^{-2}$ \\

$kT_1$ & $ 0.29\pm0.02$ keV \\
$EM_1$ & $ 7.5_{-1.5}^{+1.7}\times10^{51}$ cm$^{5}$ \\
$kT_2$ & $ 0.67_{-0.03}^{+0.04}$ keV \\
$EM_2$ & $ 11.0_{-2.6}^{+3.1}\times10^{51}$ cm$^{5}$ \\
$kT_3$ & $ 1.49_{-0.28}^{+0.61}$ keV \\
$EM_3$ & $ 5.6_{-1.6}^{+1.5}\times10^{51}$ cm$^{5}$ \\
 \hline
Element & Abundance\\
C & $ 0.9\pm0.3$  \\
N & $ 0.8\pm0.3$ \\
O & $ 0.5\pm0.2$\\
Ne & $ 1.3_{-0.1}^{+0.3} $ \\
Si & $ 0.3\pm0.1$ \\
S & $ 0.1_{-0.1}^{+0.2}  $\\
Fe & $ 0.19_{-0.04}^{+0.06} $ \\
\hline
\end{tabular}
\end{center}
\end{table}

\section{Absorption at the carbon edge}
The carbon K-edge at 43.6~\AA~ (284~eV, i.e., the energy needed to expell a K-shell electron from an isolated carbon atom) is a promising absorption feature in the \chan~ LETGS wavelength band. In this region, the line emission is relatively small compared to continuum emission. The oxygen edge, located at 23.1~\AA~ (536~eV), is also a good candidate, but with an enhanced line-to-continuum ratio. This increases the errors since the prediction of the line emission is not possible with the required precision.
Unfortunately, the UV/Ion shield of the HRC-S also contains  a large amount of carbon ($N_C \approx 10^{18}\,$cm$^{-2}$) leading to an instrumental feature very similar to that of expected interstellar/circum-stellar carbon absorption.
Still, the high continuum fraction of the first order emission around the C-K edge makes this region more promising than the O-K edge.

\subsection{The model}

As is obvious from Fig.~\ref{fig:C_trans} the height of the C edge does depend sensitively on the carbon column density (and on the grain sizes).  Therefore our basic idea to measure the height of that edge is as follows: We divide the wavelength range around the C~edge into a low \mbox{(44 -- 48~\AA)} and a high energy band \mbox{(30 -- 43~\AA)} and sum up the counts in these two bands to increase the signal. 
By comparing these two count numbers with models containing specific amounts of carbon absorption the best fit carbon column density can be found.

We do not extend the wavelength range to longer wavelengths since these longer wavelength bins are close to the detector gap at 50~\AA~ of the negative diffraction order, where the data is less reliable. We also exclude the immediate region around the edge and around the 1s-2p carbon absorption line ($\sim$44.8~\AA)
because of the relatively little known fine-structure of the carbon absorption edge. Exclusion of a larger wavelength region around the edge does not change the results significantly (see Fig.~\ref{fig:different_C_abs}).
Increasing the amount of carbon in the line of sight, changes the flux-ratio between the low and high band; since the
the change in transmission in the low energy band is comparably small, we normalise our model to match the low energy band.  Therefore
we carry out only a differential analysis.

\begin{figure}%[ht]
\centering
  \includegraphics[width=0.49\textwidth, angle=0]{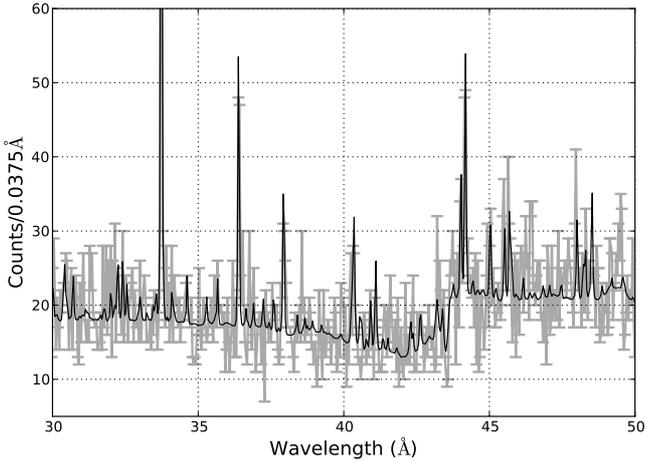}
  \caption{Model and data around the C-K edge (sum of both diffraction orders).\label{fig:Fits}}
\end{figure}

\begin{table}
\caption{
Contributions at the C-K edge (summed continuum and line emission). Listed are the total counts of the complete model (sum of positive and negative diffraction order). Top is the result from the standard light filtering and in the bottom the corresponding value for the new \texttt{Gain Map and Pulse-Height} filtering are shown. \label{tab:edge_contr}}
\begin{center}
\begin{tabular}{lrrrr}
\hline
\hline
Contribution & \multicolumn{4}{c}{Wavelength range}\\
 &\multicolumn{2}{c}{30 - 43 \AA} &\multicolumn{2}{c}{  44 - 48 \AA } \\

\hline
\multicolumn{5}{l}{\textbf{Light filter:}}\\
Total counts & 6495 && 2593 &\\
First order emission &  1546 & 23.8 \% &  1014 & 39.1 \% \\
Higher order contribution & 569 & 8.8 \% & 172 &  6.6 \%\\
Background & 4251 & 65.4 \% & 1287 & 49.6 \% \\
\hline
\multicolumn{5}{l}{\texttt{Gain Map and Pulse-Height} \textbf{filter:}}\\
Total counts & 4582 && 1939 &\\
First order emission &  1467 & 32.0 \% &  949 & 48.9 \% \\
Higher order contribution & 565 & 12.3 \% & 173 &  8.9 \%\\
Background & 2376 & 51.9 \% & 705 & 36.4 \% \\
\hline
\end{tabular}
\end{center}
\end{table}

\begin{figure}[t]
\centering
  \includegraphics[width=0.49\textwidth]{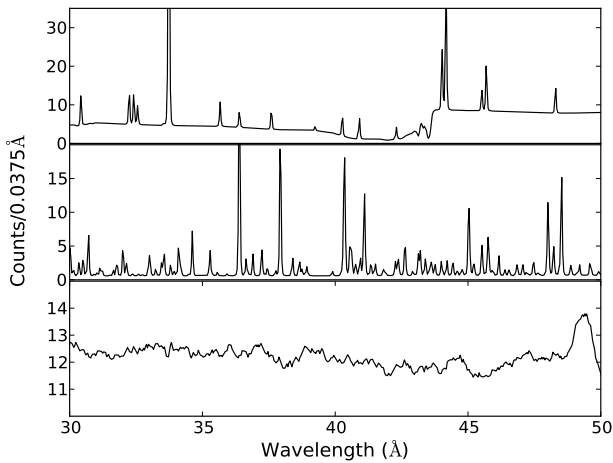}
  \caption{The different contributions to the measured counts around the C-K edge. \textbf{Top:} First order emission. \textbf{middle:} Higher order contribution. \textbf{bottom:} Background. \label{fig:edge_contributions}}
\end{figure}

For our modelling of the recorded spectrum above and below the carbon edge we consider the following
four components as contributors to the counts around the C-K edge.
\begin{enumerate}
\item The continuum emission.
\item The superimposed line emission.
\item The higher diffraction orders which cannot be filtered out due to the low intrinsic energy resolution of the photon detector (HRC-S).
\item The enhanced background of the LETGS due to a wiring error of the micro-channel plate detector.
\end{enumerate}
In the following we describe our modelling of the individual components and discuss their accuracy; the given errors pertain to the higher energy side of the edge while the numbers in brackets correspond to the lower energy side. Since the models are normalised to match the low energy side of the edge, an error at this side  will also influence the number of predicted counts at the high energy side of the edge (by roughly the same amount).
Fig.~\ref{fig:Fits} shows the complete model and the individual model contributions are shown in Fig.~\ref{fig:edge_contributions}.

The continuum is the simplest component since its shape around the C-K edge does not change noticeably for reasonable changes in the plasma 
properties.  The main spectral components are known from our global spectral modelling, but changing, for example,
the temperature by about 50~\%, changes the number of counts left of the carbon edge by less than 2~\%~(50~cts), while preserving the counts at the right side of the edge by choosing an appropriate normalisation.

The treatment of emission lines is less straightforward.  We estimate that contribution of the first diffraction order emission lines is not predictable with a better than 10~\% accuracy. Therefore, a region of 0.2~\AA~ around each strong line (based on our XSPEC/APED emission model) was excluded from the two energy bands thus removing $\sim$~98\% of the line flux of these strong lines. Only 65~\% and 73~\% of the available bins are therefore used in the 30~--~43~\AA~ and 44~--~48~\AA~ range, respectively. As
a consequence, the ratio between line and continuum emission is below 10~\% in the high and low band.  In addition, 
the remaining counts from unresolved lines are predicted from the emission model and are added to the model as a correction; these numbers turn out to be 50 and 35, respectively.
Taking this numbers as a conservative estimate of the error, the impact of these lines on the derived carbon column density is quite small, therefore, also the influence of ''unknown'' lines ought to be small.

\begin{figure}%[ht]
\centering
  \includegraphics[width=0.49\textwidth, angle=0]{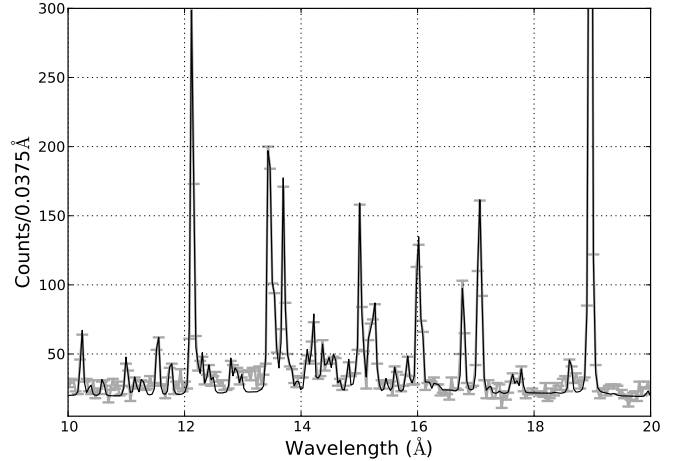}
  \caption{Data and model for the sum of the positive and the negative diffraction order. Note that the model is not based on a physical model (except for the continuum and the absorption). \label{fig:1o_model_p}}
\end{figure}

The third diffraction order of the LETGS is the strongest higher order contribution to the region around the C-K edge, 
since the second diffraction order is suppressed by the grating design. Unfortunately, \au being an active star, the Ne~IX and Ne~X 
lines are quite strong, and their third order components are easily discernible in our LETGS spectrum (cf. middle panel in Fig.~\ref{fig:edge_contributions}).
However, first and third order are always measured simultaneously, thus, by fitting the stronger (identifiable) lines independently in the wavelength range shorter than the C-K edge provides the desired description of the higher order contamination. The selection of the lines is based on the APED model (see Sect.~\ref{sect:iresults}), the \chan~HETGS spectrum of \au~ and on the lines visible in the Capella spectrum (Capella is a calibration target of the LETGS, thus providing about 400~ks of data and has an approximately comparable temperature structure as AU~Mic). Note that this method is unaffected by the uncertainties of the spectral model.  Figure~\ref{fig:1o_model_p} shows the model of the wavelength range which provides the strongest contamination at the carbon edge; 97~\% of the photons in this range are included in the model. Taking the remaining 3~\% as a measure for the accuracy of the higher order model, these 3~\% correspond to only 12 (4) photons not contained in the higher order model between 30~\AA~ and 48~\AA. We can therefore neglect this effect on the derived carbon column density compared to the other factors like the statistical noise or a potential error in the higher order diffraction efficiencies, which would for a 10~\% error \footnote{http://asc.harvard.edu/cal/} result in an error on the 30~count level (6~cts).

The last component to consider is the background. Fortunately, the background is relatively uniform in the wavelength range between 30~\AA~ and 48~\AA. Since the bins in the ``line-free'' regions are all summed up, the impact of the short-scale variation should be further reduced. The largest problem of the background in this analysis is the enlarged count number leading to an increased statistical uncertainty which amounts to about 50 (30) counts. Since the models are tuned to match the longer wavelength side of the edge, the combined statistical error is $\sim$~75 counts at the high energy side of the edge.

\subsection{The influence of carbon absorption\label{sect:nC}}

\begin{figure}[t]
\centering
  \includegraphics[width=0.49\textwidth, angle=0]{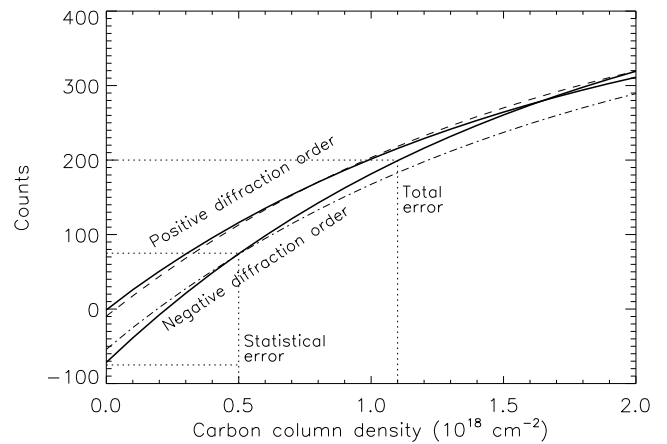}
  \caption{Number of missing (positive count number) or overpredicted photons (negative count number) as a function of the assumed carbon absorption column density in the model. The flattening of the curves indicates that the flux will, eventually, be completely absorbed in the considered wavelengths range for even higher column densities. The dashed line is the average of the positive and negative order using the new \texttt{Gain Map and Pulse-Height} filtering. The dashed-dotted line is the result using the wavelength ranges 30--42~\AA~ and 45--48~\AA.
  The statistical errors reflects only first order errors. For a discussion of the errors see Sect:~\ref{sect:nC}. \label{fig:different_C_abs}}
\end{figure}

Varying the absorbing carbon column density changes the transmission curve in the whole wavelength range (see Fig.~\ref{fig:C_trans}). 
Therefore, it is necessary to re-adjust the continuum-level and the amplitudes of the emission lines for the higher order contribution to construct a new best fit model. The influence of the carbon absorption on the low-energy side of the edge is relatively small (cf.~Fig.~\ref{fig:C_trans}),  therefore, we tuned the models so that this region matches the data by varying the continuum normalisation of the model. 
The normalisations of the models with zero absorption and $N_C=2\times10^{18}\,$cm$^{-2}$ differ by about 20~\%. Due to the strong line emission (there are not many wavelength regions without line emission at shorter wavelengths) and the refitting of these lines, this does not  noticeably change the fit-quality at the shorter wavelength. Having 
thus fixed our model to account for the number of counts on the low-energy side of the carbon edge, we can compute the model counts on the high energy side of the carbon 
edge as a function of the assumed carbon column density and hence, by comparison with the observation, the missing model counts, since increasing the carbon column density leads to a more and more relatively
reduced model count number below the edge.
The dependence of this number of missing counts in the model on the assumed carbon absorption is graphically displayed in Fig.~\ref{fig:different_C_abs}. 
Depending on the accepted accuracy of our model we can read off the maximum number of carbon atoms along the line of sight assuming gaseous absorption (i.e. no self-shielding).
If grains cause the absorption, a larger number of carbon atoms in the line of sight is required to produce the same number of missing counts. Explaining gaseous absorption with $N_C=10^{18}$~cm$^{-2}$ requires a 1.3 times higher number of carbon atoms in the line of sight for $0.1$~$\mu$m sized grains, a 1.9 times higher number for $0.3$~$\mu$m and an almost 5 times higher number for $1.0$~$\mu$m grains.

Collecting the errors attached to the individual model components,
we estimate an overall error of approximately 200 (80) counts. We cannot simply add the error of both sides of the edge, since it is rather unlikely that the same error-source results in an erroneously increased number of counts at one side of the edge while, at the same time, underestimating the counts at the other side of the edge. Therefore, we conclude that a robust estimate of the error is about 200 counts or $N_C \lesssim 10^{18}$~cm$^{-2}$ (for pure carbon gas).

The result shown in Fig.~\ref{fig:different_C_abs} suggests only a small amount of carbon-atoms along the line of sight ($N_C\sim 10^{17}\,$cm$^{-2}$), which might be supported by a slight deficit in counts around the carbon 1s-2p absorption line. However, this is probably only a statistical effect since this effect is reduced by using the stronger \texttt{Gain Map and Pulse-Height} background filtering (see dashed line in Fig.~\ref{fig:different_C_abs}).

\section{Constraining the total absorption \label{sect:nH}}

The method described above offers the ability to directly constrain the column density of carbon in the line of sight. It is, on the other hand, not useful to set tight contrains on the ``total'' absorbing column density since only a narrow wavelength range is inspected.
Emission lines located at largely different wavelengths better utilise the large wavelength coverage of the LETGS. Usage of a single element and ionisation stage reduces the dependence on the reconstruction of the abundances and the temperature structure of the corona.

A good candidate for such a study is \mbox{Fe~{\sc XVIII}} with lines at 14.21~\AA~ (blend of 14.2060 and 14.2085~\AA), 16.08~\AA~ (blend of 16.0760 and 16.0913~\AA), 93.92~\AA~ and 103.94~\AA. Assuming that all these lines are produced in the same environment, the relative fluxes of these ions depend only weakly on the temperature structure; the dependence on the density is virtually negligible for densities around $n_e \sim 10^{10}$~cm$^{-3}$ as usually found in stellar coronae \citep[e.g.][]{Ness_2002}.
Unfortunately the stronger \mbox{Fe {\sc XVIII}} lines at shorter wavelength are blended. While the 14.21~\AA~ lines do posses only small contributions from the 14.17~\AA~\mbox{Fe {\sc XXI}} line and the \mbox{Fe {\sc XVIII} doublet} at 14.26~\AA~, the 16.08~\AA~ is blended by the strong \mbox{O~{\sc VIII} doublet} at 16.00~\AA~ and a Fe~{\sc XIX} line at 16.11~\AA, which amounts to about a fourth of the \mbox{Fe {\sc XVIII}} emission for the temperatures in question. The derived fluxes of the short wavelength lines are comparable to the available HETGS data (Obs-ID 17).

\begin{figure}[t]
\centering
  \includegraphics[width=0.49\textwidth, angle=0]{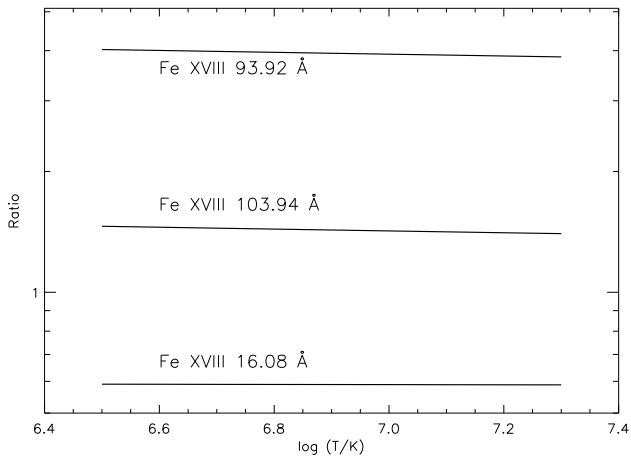}
  \caption{Ratio of the line-fluxes to the Fe~{\sc XVIII}~14.21~\AA~ line using Gaussian-like emission measure distributions with a width of $\log (T/\text{K}) = 0.5$. \label{fig:LineRatio}}
\end{figure}

The Chianti package \citep{Dere_1997, Landi_2006} was used to predict relative line fluxes. To investigate the temperature dependence of the relative fluxes, a Gaussian fit of the emission measure distribution (EMD) of Appendix~\ref{sect:DEM} was performed. The line ratios were then calculated for Gaussian EMDs with different peak temperatures and widths, e.g. the peak temperature was changed up to $\Delta \log (T) = 0.4$ (see Fig.~\ref{fig:LineRatio}).
The measured fluxes (sum of both LETGS orders) of the above lines together with predicted ratios (the 14.21~\AA~ line normalised to 1.0) are listed in Tab.~\ref{tab:nHfluxes}. 

\begin{table*}
\begin{minipage}[h]{0.99\textwidth}
\caption{Strong Fe~{\sc XVIII} lines and measured fluxes. The values for the new \texttt{Gain Map and Pulse-Height} filtering are also given. Their corresponding ratios are given in brackets in the ratio column while the values in the columns for the corresponding absorbing column densities are based on the standard filter and the value in brackets give the upper limit including the statistical error.\label{tab:nHfluxes}}

\renewcommand\footnoterule{}
\begin{tabular}{ccccccc}
\hline
\hline
 Wavelength & Measured flux & Measured flux\footnote{New \texttt{Gain Map and Pulse-Height} filter}&  Ratio to & Predicted & Corresponding & Corresponding  \\
            & \multicolumn{2}{c}{($10^{-5}$ph/cm$^2$/s)} & 14.21~\AA  & & $N_H$ (10$^{18}$~cm$^{-2}$) & $N_C$ (10$^{17}$~cm$^{-2}$) \\
\hline
14.21 & 6.3 $\pm$ 1.4 & 6.3 $\pm$ 1.4 & -- &  & \\
16.08 & 5.7 $\pm$ 1.3 & 6.3 $\pm$ 1.0 & 0.9 $\pm$ 0.3 (1.0$\pm$ 0.3) & 0.6\footnote{plus 0.2 blend}\\
93.92 & 28.9 $\pm$   3.7 & 25.8 $\pm$ 3.3 & 4.6 $\pm$ 1.1 (4.1 $\pm$ 1.1) &3.9 $\pm$ 0.3 & 0 -- 7 & 0 -- 8\\
103.94 & 11.1 $\pm$   3.4& 9.7 $\pm$2.7 & 1.8 $\pm$ 0.7 (1.5 $\pm$ 0.5)& 1.4 $\pm$0.1 & 0 -- 8 & 0 -- 10\\
\hline
\end{tabular}
\end{minipage}
\end{table*}

From the difference between predicted and measured counts the required hydrogen column density (assuming solar abundances) can be calculated. The resulting ranges are listed in Tab.~\ref{tab:nHfluxes}, they include the uncertainty in the reconstruction of the temperature and the statistical error. The influence of the adopted temperature structure is small compared to the statistical errors.
Comparison of the line fluxes from the standard filtering and that of the new \texttt{Gain Map and Pulse-Height} filtering shows that both values agree well. However, the upper limits from these fluxes are higher than that from the light filtered data since their line fluxes at longer wavelength are lower ($N_H<11\times10^{18}\,$cm$^{-2}$).

Summing the line fluxes of the short wavelengths lines and combining the upper limits of the two long wavelength lines improves the upper limit to $N_H<4\times10^{18}\,$cm$^{-2}\,(1\sigma)$.
Instead of using one single ion, we checked the ratio of the two strongest Fe lines at short and long wavelengths (\mbox{Fe {\sc XVII} 15.02~\AA} and \mbox{Fe {\sc XX} 132.85~\AA}). Their ratio depends more strongly on the temperature structure with predicted ratios (132.85~\AA/15.02~\AA) between 1.0 (log~T=6.5)and 2.0 (log~T=7.3) for a shallow EMD ($\Delta T=0.5$). However, their resulting upper limit is only slightly higher ($N_H<6\times10^{18}$~cm$^{-2}$) than that from the \mbox{Fe {\sc XVIII}} lines.

The X-ray absorption at wavelengths of around 100~\AA~ is caused mainly by He atoms. The upper limit on He from the observed line fluxes is $N_{He}<6\times10^{18}\,$cm$^{-2}$ assuming that the absorption is caused exclusively by He and $N_H<14\times10^{18}\,$cm$^{-2}$ for a pure Hydrogen absorber.
Similarly, upper limits on the abundance of other absorbing elements can be derived under the assumption that they are the only absorbing species.

The analysis of the absorption towards Capella by \citet{Gu_2006} using a comparable method has an error of about a factor of five lower which is consistent with the larger data base available for Capella which is a calibration target of the \chan~ LETGS.

\section{Discussion}
Neither the edge based method nor the line based method are able to find significant absorption along the line of sight towards AU~Mic.
For both methods, the statistical error overwhelms the error caused by the incomplete models. This is clear by inspection of Tab.~\ref{tab:nHfluxes} for the line based method but is also true for the edge analysis. The detectable carbon column density from an analysis of the edge height is $5\dots50\times10^{17}\;$cm$^{-2}$ for artificial stars with X-ray fluxes within a range of five around that of AU~Mic (50~ks LETGS exposure). This value decreases only slowly for even higher X-ray fluxes and ``saturates'' at $N_C\approx3\times10^{17}\;$cm$^{-2}$ due to the potential error in the higher order contribution (see Appendix~\ref{sect:Cdetectability}).
The edge-method provides a direct
upper limit on the carbon content of the disk without any assumption on the absorber. Its limit is about that provided by the line based method assuming that only carbon is in disk, i.e., calculating the maximum carbon column density allowing the detection of the line-flux from the long wavelengths \mbox{Fe \sc{XVIII}} lines.

Two measurements of H (atomic and molecular) restrict the total hydrogen column density in the line of sight towards AU~Mic to $N_H\lesssim 4\times10^{19}\,$cm$^{-2}$ \citep[mainly $H_2$,][]{Roberge_2005}. From this value,  only $2\times10^{18}\,$cm$^{-2}$ are atomic hydrogen \citep{Wood_2005}. Note that the conversion of molecular hydrogen to $N_H$ in stellar disks is uncertain \citep[e.g.,][]{Kamp_2007}, while molecular hydrogen is orders of magnitude less abundant than atomic hydrogen in the nearby interstellar medium \citep[$d<200$~pc;][]{Lehner_2003}.
The location of the fluorescent $H_2$ detected by \citet{France_2007} is not necessarily along the line of sight and leaves space for an additional $H_2$ not contributing to the fluorescent $H_2$ emission.

X-rays are sensitive to both, atomic and molecular, hydrogen. $H_2$ absorption is about 2.8 times stronger than that of a single $H$ atom \citep{Wilms_2000}. Furthermore, the X-ray absorption is insensitive to the excitation state of $H_2$. Therefore, the 5$\sigma$ upper limit from the UV measurement ($N_{H_2}<17\times10^{18}\,$cm$^{-2}$) corresponds to about $N_H\lesssim50\times10^{18}\,$cm$^{-2}$ in soft X-rays; the same order as the upper limit derived from the line fluxes assuming a pure hydrogen absorber.
The X-ray upper limit on $H$ also includes $H_2$ in the line of sight located in the inner and outer part of the disk which might have different excitation stages and therefore complements the upper limit of \citet{Roberge_2005}. Furthermore, the detected line fluxes at $\lambda\gtrsim100\,$\AA~ can be translated to upper limits on other elements restricting the gas / small grain content of the disk, e.g., that of carbon in gaseous form and small grains ($s\lesssim 0.3~\mu\mbox{m}$, see Tab.~\ref{tab:nHfluxes}).

\section{Summary}

We analysed the impact of absorption caused by AU~Mic's debris disk on its observed X-ray spectrum resulting in three upper limits on the column densities for hydrogen, helium and carbon:
\begin{itemize}
\item $N_H<2.8\times10^{19}\,$cm$^{-2}$ (pure H absorption)
\item $N_{He}<1.2\times10^{18}\,$cm$^{-2}$ (pure He absorption)
\item $N_C<10^{18}\,$cm$^{-2}$ (pure C absorption)
\end{itemize}
Assuming an absorber with solar abundances, the upper limit is $N_H<10^{19}$~cm$^{-2}$, which is about a factor of five higher than the pure interstellar value.
The statistical error caused by counting statistics and the high background prevents to set lower limits, while the presented methods are in principle more sensitive.

Both upper limits are consistent with the idea that the AU~Mic disk is optically thin in the radial direction as proposed by recent analyses from optical and infrared measurements, and in line with current disk models, which assume that the inner part of the disk is almost void of small grains. The debris of the collisions of planetesimals in the ``birth-ring'' populate mainly the outer part of the disk, where their density is so low that they escape a detection in this X-ray observation. In the inner part of the disk larger grains hold the mass as predicted by birth-ring scenarios.

\citet{Roberge_2005} state that they find weak signs of $H_2$ absorption on the order of $10^{18}\,$cm$^{-2}$ in their UV data. 
To reach this level with the \chan~LETGS setup, longer integration times are required. For Capella, a calibration target of the LETGS, the $N_H \sim 10^{18}$~cm$^{-2}$ range is accessable. However, it seems unlikely to reach down to $N_H \sim 10^{17}$~cm$^{-2}$ with the present instrumentation as would be required to safely distinguish a hypothetical circum-stellar $H_2$ component in the $N_H\sim10^{18}\,$cm$^{-2}$ range from the interstellar atomic $H$ absorption.

\appendix
\section{Higher order response matrices for XSPEC\label{sect:RespHigherOrders}}
The response matrices contain the probabilities describing how the detector will respond on a photon with a given energy.
Including the higher diffraction orders therefore requires to add additional probabilities at two, or more times the rest wavelength. Since the resolution grows with the order number, the energy grid needs to be refined accordingly.
The FITS-specification for response files (CAL/GEN/92-002) already includes the concept of wavelength groups which are ideal to reduce the size of the final response matrix.

\section{Carbon detectability\label{sect:Cdetectability}}
\begin{figure}[ht]
\centering
  \includegraphics[width=0.49\textwidth, angle=0]{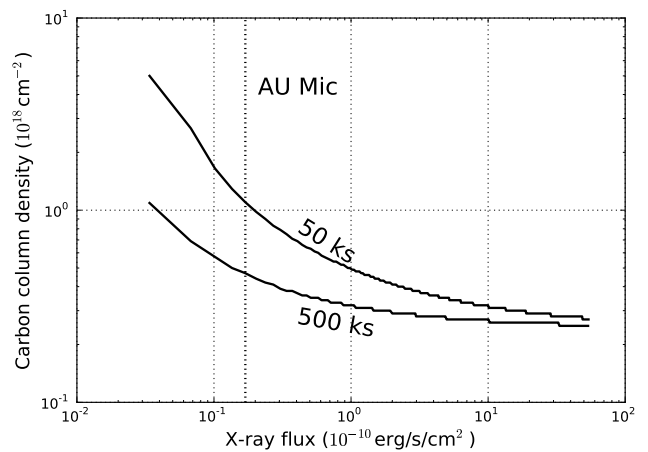}
  \caption{The detectable carbon column density by measuring the jump height at the carbon edge as a function of the X-ray flux (0.1~keV -- 2.0~keV). Here, a pure carbon disk is assumed. The two lines represent a 50~ks and a 500~ks exposure with the LETGS. The same temperature structure as AU~Mic is assumed which determines the higher order contribution around the carbon edge. The vertical dotted line indicates the X-ray flux of AU~Mic.\label{fig:detecta}}
\end{figure}

To estimate the detectability of carbon absorption from the edge height, we need to quatify the different error contributions as a function of the source count-rate. Concerning the error due to unknown unresolved first order emission lines and uncertainties in the temperature structure, we regard a correlation with the statistical error as realistic, since better data quality enables a more precise temperature reconstruction and, in turn, a better model for the emission lines around the carbon edge. However, the detailed dependence on the source count-rate might be more complicated.
The higher order contribution scales with the source flux; therefore, its error scales also linearly with the first order flux since calibration uncertainties dominate the error. Figure~\ref{fig:detecta} shows the minimum detectable carbon column density for different source fluxes assuming a pure carbon disk. It shows that column densities below $3\times10^{17}\;$cm$^{-2}$ can only be achieved for star an order of magnitude X-ray brighter than AU~Mic and with a hypothetical 500~ks observation. However, the existence of other elements in the disk increases the detectable column density due to the absorption of the first order flux around the carbon edge. Furthermore, without knowing the composition of the disk, the extra absorption changes the continuum slope around the carbon edge and therefore increased the uncertainty in the carbon column density.

\section{The emission measure distribution \label{sect:DEM}}
The large wavelength coverage of the LETGS offers the ability to quantify the emission measure distribution (EMD) in the corona. Using the lines listed in Tab.~\ref{tab:lines}, we reconstructed the EMD using the PintOfAle \citep{Kashyap_2000} package. Our result compares well the result obtained using the line fluxes measured in the UV wavelength range by \citet{DelZanna_2002}.
We also experimented with the Chianti package \citep{Dere_1997, Landi_2006} and found that the results are strongly influenced by numerical problems due to the shallow temperature coverage of the observed lines. However, both methods give rather similar results.

\begin{figure}[ht]
\centering
  \includegraphics[width=0.49\textwidth, angle=0]{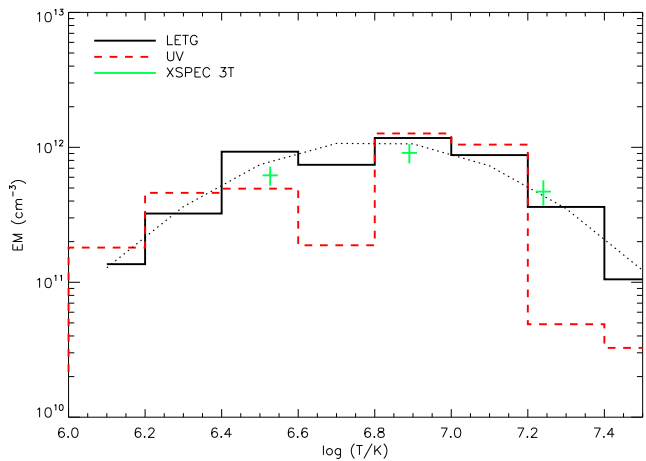}
  \caption{The reconstructed emission measure distribution. The dotted line indicates the Gaussian fit to the EMD. \label{fig:EMD}}
\end{figure}

It is clear, that most of the emission is produced in the temperature range $6.4 < \log (T / \text{K} ) < 7.2$. We also show in Fig.~\ref{fig:EMD} the location of the three temperature components from the XSPEC fit with APEC models (variable abundances). For an estimate of the error in the relative linefluxes in Sect.~\ref{sect:nH} we fitted the emission measure distribution with a single Gaussian. We can then change the centroid and the width of the Gaussian to analyse the impact of an error in the reconstructed temperature distribution on the relative linefluxes.

\begin{table}
\caption{Strong lines and measured fluxes.\label{tab:lines}}
\begin{center}
\begin{tabular}{lrr}
\hline\hline
 Ions & Wavelength & Flux (10$^{-5}$ph/cm$^2$/s) \\
 \hline
 Si XIII &  6.65 &  4.5 $\pm$   0.7 \\
    Ne X & 10.24 &  6.3 $\pm$   1.0 \\
   Ne IX & 11.54 &  6.3 $\pm$   1.0 \\
    Ne X & 12.13 & 44.1 $\pm$   2.1 \\
   Ne IX & 13.45 & 32.8 $\pm$   1.9 \\
   Ne IX & 13.54 & 11.2 $\pm$   1.3 \\
   Ne IX & 13.70 & 22.9 $\pm$   1.6 \\
Fe XVIII & 14.21 &  6.3 $\pm$   1.4 \\
 Fe XVII & 15.01 & 18.0 $\pm$   1.4 \\
Fe XIX/O VIII & 15.18 &  6.4 $\pm$   1.1 \\
Fe XVII/Fe XX  & 15.26 &  8.4 $\pm$   1.1 \\
  O VIII & 16.01 & 17.6 $\pm$   1.5 \\
Fe XVIII & 16.08 &  5.7 $\pm$   1.3 \\
 Fe XVII & 16.78 & 12.1 $\pm$   1.2 \\
Fe XVII  & 17.10 & 30.2 $\pm$   1.9 \\
  O VIII & 18.97 & 110.3 $\pm$   3.3 \\
   O VII & 21.61 & 29.2 $\pm$   2.3 \\
   O VII & 21.79 &  10.0 $\pm$   1.6 \\
   O VII & 22.10 & 19.4 $\pm$   2.0 \\
   N VII & 24.79 & 16.6 $\pm$   2.0 \\
    C VI & 33.73 & 33.9 $\pm$   2.9 \\
Fe XVIII & 93.89 & 28.9 $\pm$   3.7 \\
Fe XVIII & 103.92 & 11.1 $\pm$   3.4 \\
  Fe XIX & 108.33 & 26.2 $\pm$   4.3 \\
   Fe XX & 118.66 & 14.1 $\pm$   4.2 \\
Fe XX & 121.85 & 17.0 $\pm$   5.0 \\
Fe XXI & 128.75 &  $<$16.2 \\
   Fe XX & 132.85 & 37.7 $\pm$   8.6 \\
 Fe XXII & 135.78 & 27.6 $\pm$   7.9 \\

\hline
\end{tabular}
\end{center}
\end{table}

\acknowledgement{
We thank Dr. P. Predehl (MPE) for kindly providing the opportunity to observe AU~Mic with the {\it Chandra}~LETGS.
This work has made use of data obtained from the \chan~ and \xmm~ data archives.
CHIANTI is a collaborative project involving the NRL (USA), RAL (UK), MSSL (UK), the Universities of Florence (Italy) and Cambridge (UK), and George Mason University (USA). PCS acknowledges support from the DLR under grant 50OR703.}

\bibliography{au}
\end{document}